\documentclass[12pt]{iopart}

\usepackage{graphicx}% Include figure files
\usepackage{dcolumn}% Align table columns on decimal point
\usepackage{bm}% bold math
\usepackage[left]{lineno}
\usepackage{gensymb}
\begin{document}

\title{On the jets emitted by driven Bose--Einstein condensates}% Force line breaks with \\
\author{Miguel Arratia}
\address{Department of Physics, University of California, Berkeley, CA 94720, USA.}%
\ead{marratia@berkeley.edu}
%\affiliation{Nuclear Science Division, Lawrence Berkeley National Laboratory, Berkeley, CA 94704, USA}
\date{\today}% It is always \today, today,
             %  but any date may be explicitly specified
%%%%%%%%%%%%%%%%%%%%%%%%%%%%%%%%%%%%%%%%%%%%%%%%%%%%%%%%%%%%%%%%%%%%%%%%%%%%%%%%%%%%%%%%%%%%%%
\begin{abstract}
Features of the emission of jets by driven {Bose--Einstein} condensates, discovered by {Clark et al. (Nature 551, 356359)}, can be understood by drawing analogies with particle and nuclear physics. In particular, the widening of the {$\Delta\phi=\pi$} peak in the angular correlation function is due to a dijet acollinearity, which I estimate to be about 5\degree~RMS. I also propose new correlation studies using observables commonly used in studies of the quark-gluon plasma. 
\end{abstract}
%%%%%%%%%%%%%%%%%%%%%%%%%%%%%%%%%%%%%%%%%%%%%%%%%%%%%%%%%%%%%%%%%%%%%%%%%%%%
\pacs{Valid PACS appear here}% PACS, the Physics and Astronomy
                             % Classification Scheme.
%\keywords{Suggested keywords}%Use showkeys class option if keyword
                              %display desired
\maketitle
%\tableofcontents
\section{\label{sec:level1}Introduction}
%%%%%%%%%%%%%%%%%%%%%%%%%%%%%%%%%%%%%%%%%%%%%%%%%%%%%%%%%%%%%%%%%%%%%%%%%%%%%%%%%%%
Clark et al.~\cite{Clark} recently discovered a new phenomenon in which a stimulated Bose--Einstein condensate emits a burst of collimated jets of atoms. The data analyzed by means of a second-order angular correlation function show two peaks at {$\Delta\phi=0$} and {$\Delta\phi=\pi$}. These were attributed to back-to-back emission of jets, reflecting momentum conservation in the primary atom--atom scattering that triggers the dijet runaway formation. 

However, their data deviates from the predicted correlation, specially in the region of the the peak at {$\Delta\phi=\pi$}. This is much wider than the peak at {$\Delta\phi=0$} and has only about 70--85\% of its integral. This fact is not totally understood, and the authors claimed in Ref.~\cite{Clark} that {\it ``further investigation into the differences between the two peaks is required''}. 

%The angular spread of the jets, measured with the {$\Delta\phi=0$} peak in the correlation function, was shown to arise from the Heisenberg uncertainty principle due to the confinement of atoms in the Bose--Einstein condensate. 
Here, I offer an explanation for the broadening of the peak at {$\Delta\phi=\pi$} by drawing analogies with observations in high-energy particle and nuclear physics. I also propose new measurements on this jet phenomenon inspired in studies of the quark--gluon plasma. 

This article is organized as follows: section~\ref{sec:acoplanarity} shows an estimate of the dijet acollinearity present in the Clark et al. experiment by drawing an analogy with proton--proton collisions with ``intrinsic parton $k_{T}$''; section~\ref{sec:vertical} discusses 
the vertical dimension of the Bose--Einstein condensate; section~\ref{sec:proposals} shows proposals of new observables; and section~\ref{sec:conclusions} describes the conclusions.
%%%%%%%%%%%%%%%%%%%%%%%%%%%%%%%%%%%%%%%%%%%%%%%%%%%%%%%%%%%%%%%%%%%%%%%%%%%%%%%%%%%%%%%%%%%
\section{\label{sec:acoplanarity}Dijet acollinearity}
\subsection{Parton--parton scattering}
In the mid seventies, the production of roughly back-to-back sprays of collimated hadrons (jets) in proton collisions was attributed to collinear parton--parton scattering with large momentum transfer. However, this model did fail to describe the data from experiments at the CERN Intersecting Storage Rings---the world's first hadron collider. 

In 1977, Feynman et al.~\cite{Feynman} modified the collinear parton--parton scattering by introducing an {``extra kick''} to the partons,  intrinsic parton $k_{T}$, that yielded a dijet acollinearity. This allowed them to explain, among other things, the data from two-particle correlations that showed a peak at {$\Delta\phi=\pi$} that was broader than the peak at {$\Delta\phi=0$}. 
%To this day, this effect is modeled in the complex simulations widely used by the particle and nuclear physics community~\cite{Pythia1}. Its origin is attributed to initial-state gluon emission (QCD next-to-leading order effects).  
%%%%%%%%%%%%%%%%%%%%%%%%%%%%%%%%%%%%%%%%%%%%%%%%%%%%%%%%%%%%%%%%%%%%%%%%%%%%%%%%%%%%%%%%%%%%%%%%%%%%%
\subsection{Atom--atom scattering}
Clark et al. compared their measured second-order angular correlation function, $g^{2}$, with an analytically calculation given by: 
\begin{equation}
\label{corr}
g^{2}(\Delta\phi) = 1 + \left|\frac{2J_{1}(k_{f}R\Delta\phi)}{k_{f}R\Delta\phi}\right|^{2}+\left|\frac{2J_{1}(k_{f}R[\Delta\phi-\pi])}{k_{f}R[\Delta\phi-\pi]}\right|^{2},
\end{equation}
where $J_{1}$ is the first Bessel function (resulting from the Fourier transform of the density of a {two-dimensional} uniform disk), $k_{f}$ is the wavenumber of the ejected atoms, and $R$ is the radius of the {Bose--Einstein} condensate. 

This function shows two identical peaks at {$\Delta\phi=0$} and {$\Delta\phi=\pi$}, reflecting the assumption of exactly back-to-back emission of jets that is based on {\it ``conservation of momentum in the underlying pair-scattering process"}\cite{Clark}. I suggest that the deviation of data from equation~\ref{corr} arises from a small dijet acollinearity.% that reflects an acollinearity in the primary atom--atom scattering. 

\subsection{Estimate of dijet acollinearity}
The dijet acollinearity can be estimated from the widths of  {$\Delta\phi=0$} and {$\Delta\phi=\pi$} peaks in the measured $g^{2}(\Delta\phi)$, following a method first used in particle physics by the CCOR collaboration~\cite{CCORS} about 40 years ago, and more recently by the PHENIX collaboration~\cite{PHENIX,JJ}. This method relies on a Gaussian approximation for the jet transverse spread and basic trigonometry to obtain an average angle of dijet acollinearity. 

For the case of the jets observed in the Clark. et al. experiment, the equations involved are simplified because all the jet constituents (atoms) have roughly the same momentum instead of being power-law distributed like in hadrons of QCD jets. In the Clark et al. experiment, the average atom transverse momentum relative to the jet axis is the same for all jets. This is because they are the result of a bosonic enhacement and their angular spread reflects the size of the source (i.e. a Handbudy Brown and Twiss bunching). The average transverse momentum can be estimated from the {$\Delta\phi=0$} peak of the correlation function, $\sigma_{\mathrm{N}}$. This can be combined with the width of the {$\Delta\phi=\pi$} peak, $\sigma_{\mathrm{A}}$, to extract the average dijet acollinearity:

\begin{equation}
\langle \phi\rangle \approx \frac{\langle k_{T}\rangle}{k_{f}} 
= \frac{1}{\sqrt{2}}\sqrt{\sin^{2}\left(\sqrt{2} \frac{\sigma_{A}}{\sqrt{\pi}}\right) - \large(\frac{\sigma_{\mathrm{N}}}{\sqrt{\pi}}\large)^{2}}.
\label{eq:kt}
\end{equation}
From Ref.~\cite{Clark}, we know that the half-maximum half-width of the {$\Delta\phi=0$} peak is about {2\degree}, for {$R$=8.5~$\mu$m} and {$f$=2 kHz}, and the width of the {$\Delta\phi=\pi$} peak is about three times larger. It follows from Equation~\ref{eq:kt} that $\langle \phi\rangle$ is about of about 5\degree.
%\begin{equation}
%\phi_{0} \approx \arcsin (\sqrt{2}\langle k_{Ty}\rangle/k_{f}) \approx 4\degree.
%\label{eq:acolinearity}
%\end{equation}
Note that this small angle has a big effect in the width of the {$\Delta\phi=\pi$} peak (that measures inter-jet correlations) but no effect in the {$\Delta\phi=0$} peak (that measures intra-jet correlations).   

\subsection{Numerical calculation of $g^{2}$ with dijet acollinearity}
To illustrate the effect of a dijet acollinearity on the measured $g^{2}$ function, I used a simple numerical simulation in which the azimuthal angle between the jet centers is drawn from a Gaussian with standard deviation of 5\degree~; the jets angular density, $n(\phi)$, is approximated by a Gaussian with a standard deviation of 1.5\degree. The $g^{2}$ function is calculated as:

\begin{equation}
g^{2}(\Delta\phi) = \frac{\langle\int d\theta n(\theta)n(\theta+\Delta\phi)\rangle}{\langle \int d\theta n(\theta)\rangle^{2}},
\end{equation}
where the average is taken over 1000 draws of different acollinearity angles, minimizing the statistical uncertainty on the calculation. 

\begin{figure}
\centering
\includegraphics[width=0.78\columnwidth]{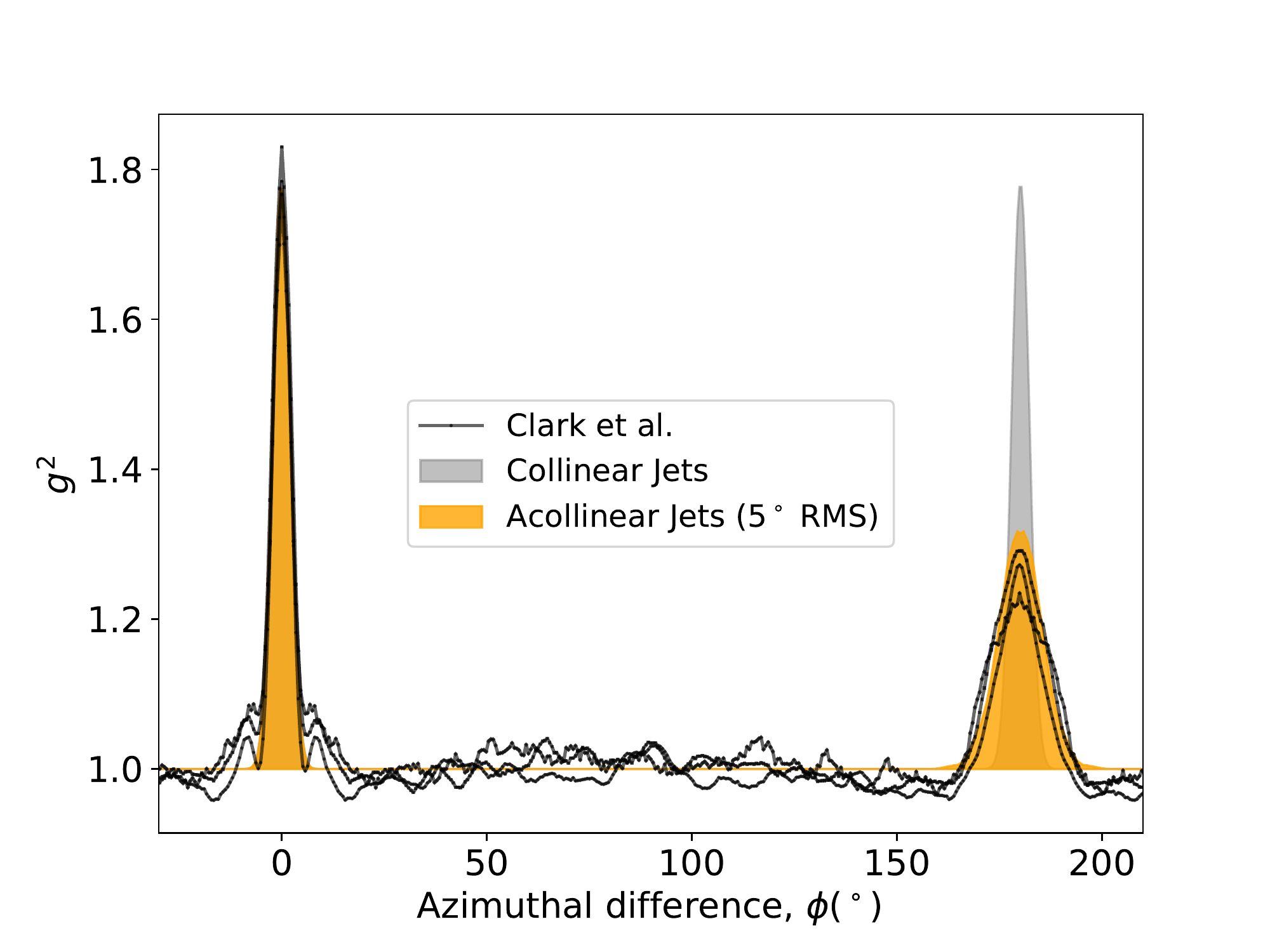}
\caption{Angular correlation function, $g^{2}$, from Ref~\cite{Clark} (black) and numerical calculations assuming back-to-back dijets (gray) and acollinear dijets with an acollinearity of 5\degree~on RMS (orange).}
\label{simulation}
\end{figure}

Figure~\ref{simulation} shows the result of this calculation with data from the Clark et al. experiment. The calculated $g^{2}$ is multiplied by a constant factor such that the height of the $\Delta\phi=0$ peak roughly matches the data. The width of the $\Delta\phi=0$ peak in the calculation matches the width of the measured peak by construction; the discrepancy in the tails arise due to the Gaussian approximation of the jet densities. The width of the $\Delta\phi=\phi$ peak of the calculation matches the data well. 

This result is based on general assumptions and it simply states that the data in Ref.~\cite{Clark} can be explained with a dijet acollinearity of $5\degree$ RMS. The explanation of the origin of such dijet acollinearity lies beyond the scope of this work. In Ref.~\cite{AngularMomentum}, this was attributed to {\it``the destructive interferences between atoms with different angular momenta''}.

%What is the origin of this dijet acollinearity? I argue that it might reflect an acollinearity in the original atom--atom scattering. Thermal motion cannot cause momentum imbalance because the temperature involved is extremely low (7 nK). However, a lower limit for this atom--atom acollinearity is imposed by the Heisenberg uncertainty principle, which sets $\Delta k\approx 1/R$ for all atoms in the Bose-Einstien. Thus, the scattered atoms should form a minimum angle of about {$\phi\approx {\sqrt{2}/Rk_{f}}$ $\approx 2\degree$} (for {$R$=8.5~$\mu$m}, and {$f$=2 kHz}). Here the factor $\sqrt{2}$ accounts for the random sum of momenta. 

%The estimated lower limit for acollinearity is very close to the value estimated from data. However, I note that other dynamics of the driven Bose-Einstein condensate might also play a role.
\section{\label{sec:vertical}Vertical direction}
%%%%%%%%%%%%%%%%%%%%%%%%%%%%%%%%%%%%%%%%%%%%%%%%%%%%%%%%%%%%%%%%%%%%%%%%%%%%%%%%%%%%%%%%%%%%%%%%%%%%%%%%%%%%%%%%%
The dimensions of the Bose--Einstein condensate described in Ref.~\cite{Clark} are given by a typical $R$ value of 8.5~$\mu$m, and vertical extend of 0.5 $\mu$m (root-mean-square). So, most of the atoms scattered with a polar angle smaller than {$\arcsin(0.5/8.5)\approx3 \degree$} will traverse most of the {Bose--Einstein} condensate, thus forming an observable dijet. 

As noted by Clark et al., some atoms within a jet might lie outside the field-of-view of the experiment (in particle physics jargon, this is an acceptance loss due to limited pseudorapidity coverage). This can explain part of the 15--30$\%$ difference of the integral of the peaks at {$\Delta\phi=0$} and {$\Delta\phi=\pi$}, and the discrepancy between equation~\ref{corr} and data at the {$\Delta\phi=0$} peak. 

Here, I suggest that this loss could be corrected, or at least be considered when calculating the predicted correlation function. This should consider the vertical structure of the Bose--Einstein condensate. The improved calculation might describe data better and serve as a baseline to search for anomalous effects, such as those described in Ref.~\cite{Ogren} and Section~\ref{sec:proposals}.
%\section{\label{sec:origins} On the origin of the jets}
%Clark et al attribute 
%%%%%%%%%%%%%%%%%%%%%%%%%%%%%%%%%%%%%%%%%%%%%%%%%%%%%%%%%%%%%%%%%%%%%%%%%%%%%%%%%%%%%%%%%%%%%%%%%%%%%%%%%%%%%%%%%%%%%%%%%%%%%%%%%%%%%%%%%%
\section{\label{sec:proposals}Proposed new studies}
Clark et al. suggested that {\it ``one could probe excitations that are present in more exotic states of matter by amplifying them to form detectable jets"}~\cite{Clark}. That would not be the first time that ``jets" are used as ``probes" of exotic states of matter. Here I suggest measurements inspired in angular correlations and jet studies that probe the quark--gluon plasma, which is also a strongly interacting system.

The events shown in Ref.~\cite{Clark} have multiple dijets. The authors claim that the dijet directions are random. However, I note that the spacing between dijets looks suspiciously uniform. Given that driven Bose-Einstein condensates are a quantum many-body system, is not unreasonable to expect an overall pattern caused by a collective behaviour. This might be even more evident when probing the appearance of vortices, solitons, and other exotic effects alluded in Ref.~\cite{Clark}. 

To further study this and search for more complex correlations than what is caused by momentum conservation and HBT bunching, I suggest to perform a multi-particle correlation study like the ones described in Refs.~\cite{Cumulant1,Cumulant2,Cumulant3}. These ``cumulants'' techniques were designed to study the collective behaviour caused by hydrodynamical flow of the quark-gluon plasma, which manifests as an anisotropy in the particle emission. These techniques suppress ``non-flow'' correlations that arise due to momentum conservation, jets, and HBT correlations. 

While in principle these sources of correlations can be suppressed using higher-order correlation functions, $g^{n}$ with large $n$, in practice the calculations get cumbsersome quickly. In contrast the cumulants analysis can use all particles in the event in an efficient way. More importantly, they can reveal true collective behaviour that might be obscured by strong correlations among a small number of atoms\footnote{After the release of the first draft of this manuscript, the Chicago group independently released a preprint with a measurement of higher-order correlations~\cite{ComplexCorrelations}; they indeed observed higher-order correlations indicating a more complex pattern than reported in~\cite{Clark}}. 
%%%%%%%%%%%%%%%%%%%%%%%%%%%%%%%%%%%%%%%%%%%%%%%%%%%%%%%%%%%%%%%%%%%%%%%%%%%%%%%%%%%%%%%%%%%%%%%%%%%%%%%%%%%%%%%%%%%%%%%%%%%%%%%%%%%%%%%%%%%%%%%%%%%%%%%%%%%%%%%%%%%%%%%%%%%%%%%%%%%%%%%%%%%%%
\section{\label{sec:conclusions}Conclusions}
In conclusion, this work explains features of the novel phenomenon of atom jets emitted by driven Bose--Einstein condensates. The broader peak in the angular correlation function at {$\Delta\phi=\pi$} can be explained by a dijet acollinearity of about {5\degree}. I also have suggested novel observables for the study of driven Bose--Einstein condensates inspired by studies of the quark-gluon plasma. This is among first papers on the phenomenology of jets in driven Bose--Einstein condensates.
%%%%%%%%%%%%%%%%%%%%%%%%%%%%%%%%%%%%%%%%%%%%%%%%%%%%%%%%%%%%%%%%%%%%%%%%%%%%%%%%%%%%%%%%%%%%%%%%%%%%%%%%%%%%%%%%%%%%%%%%%%%%%%%%%%%%%%%%%%%%%%%%%%%%%%%%%%%%%%%%%%%%%%%%%%%%%%%%%%%%%%%%%%%%%%%%%%%%%%%%%%%%%%%%%%%%%%%%%%%%%%%%%

\section{\label{sec:level1}Acknowledgements}
I thank the Chicago group for providing the data of Ref.~\cite{Clark} and useful discussions, and to the LBNL RNC group for useful discussions.
%I am grateful to the members of the Relativistic Nuclear Collisions at Lawrence Berkeley National Laboratory for discussions about this work.
%I am grateful to Stanley Brodsky, Boris Kopeliovich, Marco van Leeuwen, Nu Xu, Barbara Jacak, Yue Shi Lai, Peter Jacobs, Brian Cole, and Logan Clark for discussions about this work. I am grateful to Jacqueline Garrido for checking this manuscript. 


\section*{References}

\begin{thebibliography}{9}
\bibitem{Clark}
L. W. Clark et al.,
%\textit{``Collective emission of matter-wave jets from driven Bose--Einstein condensates"},
Nature 551, 356–359.
\bibitem{Feynman}
R. P. Feynman et al.,
%\textit{``Correlations Among Particles and Jets Produced with Large Transverse Momenta"}, 
Nucl.~Phys~ B128 (1977) 1-65.
\bibitem{Pythia1}
T. Sjostrand et al.,
%\textit{``PYTHIA 6.4 Physics and Manual"},
JHEP 0605:026,2006.
\bibitem{CCORS}
CCOR Collaboration, 
%\textit{``A Study of Final States Containing High-$p_{T}$ $\pi^{0}$s atthe CERN ISR''},
1979 Phys.~Scr. 19 116.
\bibitem{PHENIX}
PHENIX Collaboration,
%\textit{``Jet Properties from Dihadron Correlations in p+p Collisions at $\sqrt{s}=$200 GeV"},
Phys.~Rev.~D74:072002,2006.
\bibitem{JJ}
J. Jia, 
%\textit{``Probing jet properties via two particle correlation method"},
J.~Phys.~G31 (2005) S521-S532.
\bibitem{Ogren}
M. Ogren, K. V. Kheruntsyan,
%\textit{``Atom-atom correlations in colliding Bose--Einstein condensates"},
Phys.~Rev.~A 79, 021606(R) (2009).
\bibitem{AngularMomentum}
Z. Wu, H. Zhai, 
arXiv:1804.08251v2.
\bibitem{ComplexCorrelations}
L. Feng et al.,
arXiv:1803.01786.



%\bibitem{Farrar}
%G. Farrar et al.,  
%\textit{Transparency in nuclear quasiexclusive processes with large momentum transfer},
%Phys. Rev. Lett. 61, 686 (1988).
%\bibitem{ColorTransparency}
%D. Dutt et al.,
%\textit{``Color Transparency: past, present and future''},
%Prog.~Part.~Nucl.~Phys. 69 (2013) 1-27.
%\bibitem{ColorBoris}
%B.~Z.~Kopeliovich, et al., 
%\textit{``Quenching of high-pT hadrons: Energy Loss vs Color Transparency,''}
%Phys.~Rev.~C  86, 054904 (2012)
%\bibitem{tomography}
%Burke, Karen M. et al.,
%\textit{``Extracting the jet transport coefficient from jet quenching in high-energy heavy-ion collisions''},
%Phys.~Rev.~C90 (2014) no.1, 014909.
\bibitem{Cumulant1}
A. Bilandzic et al.,
%\textit{``Flow analysis with cumulants: Direct calculations"},
Phys.~Rev.~C 83 (2011) 044913.
\bibitem{Cumulant2}
R. S. Bhalerao et al.,
%\textit{``Analysis of anisotropic flow with Lee-Yang zeroes"},
Nucl.~Phys. A 727 (2003) 373.
\bibitem{Cumulant3}
N. Borghini et al.,
%\textit{``Anisotropic flow from Lee-Yang zeroes: A practical guide"},
J. Phys. G 30 (2004) S1213.
%\bibitem{Tannenbaum}
%M. J. Tannenbaum,
%\textit{``Measurement of $\hat{q}$ in Relativistic Heavy Ion Collisions using di-hadron correlations''},
%Physics Letters B771(2017)553-557.
%\bibitem{FermiGas}
%K. M. O'Hara et al., 
%\textit{Observation of a Strongly Interacting Degenerate Fermi Gas of Atoms},
%Science 298 2179 (2002).
%\bibitem{FermiGasII}
%B. Clancy, L. Luo, and J. E. Thomas,
%\textit{Observation of Nearly Perfect Irrotational Flow in Normal and Superfluid Strongly Interacting Fermi Gases}
%Phys. Rev. Lett. 99, 140401 (2007). 



\end{thebibliography}
\end{document}